\documentclass[11pt,onecolumn]{IEEEtran}

\usepackage{amsmath,amssymb,bm,cite,graphicx,texdraw, epsfig}
\newcommand{\beq}{\begin{equation}}
\newcommand{\eeq}{\end{equation}}
\newcommand{\bea}{\begin{eqnarray}}
\newcommand{\eea}{\end{eqnarray}}
\newcommand{\bef}{\begin{figure}}
\newcommand{\eef}{\end{figure}}
\newcommand{\bd}{\begin{displaymath}}
\newcommand{\ed}{\end{displaymath}}
\newcommand{\nn}{\nonumber}
\newcommand{\nnl}{\nonumber \\}
\newcommand{\fig}[1]{Fig.\ \ref{#1}}

\def\bmat{\left[ \begin{array}}
\def\emat{\end{array} \right]}
\newcommand{\defn}{\stackrel{\triangle}{=}}

\newtheorem{theo}{Theorem}

\begin{document}
\bibliographystyle{ieeetr}

\title{Robust hypothesis testing with a relative entropy
tolerance}
\author{Bernard C. Levy\thanks{
B. C. Levy is with the Department of Electrical and Computer
Engineering, University of California, Davis, CA 95616, USA 
(e-mail: levy@ece.ucdavis.edu).}}

\date{\today}

\maketitle

\IEEEaftertitletext{\vspace*{-0.2in}}

\begin{abstract}

This paper considers the design of a minimax test for two
hypotheses where the actual probability densities of the
observations are located in neighborhoods obtained by
placing a bound on the relative entropy between actual
and nominal densities. The minimax problem admits a saddle
point which is characterized. The robust test applies a
nonlinear transformation which flattens the nominal 
likelihood ratio in the vicinity of one. Results
are illustrated by considering the transmission of 
binary data in the presence of additive noise.

\end{abstract}

\begin{keywords}

Robust hypothesis testing, Kullback-Leibler divergence,
min-max problem, saddle point, least favorable densities.

\end{keywords}

\section{Introduction}
\label{sec:intr}

Robust hypothesis testing and signal detection problems have
been examined in detail over the last 40 years \cite{Hub3,KP}.
The purpose of such studies is to design tests or 
detectors which are insensitive to modelling errors.
Specifically, whereas standard Bayesian or Neyman-Pearson
tests are designed for nominal observation probability 
distributions, their performance may degrade rapidly
when the actual model deviates only moderately
from the nominal model. To guard against modelling
errors, a minimax framework is usually adopted for 
selecting tests or detectors. In this context, the
goal is to design a test that minimizes the  
worst-case performance for all observation models
in a properly specified neighborhood of the nominal   
model. For robust hypothesis testing, when the
neighborhood of the nominal model under each hypothesis 
corresponds either to a contamination model or
a proximity model based on the Kolmogorov metric or a
variant thereof, Huber \cite{Hub1,Hub2,Hub3} showed that 
the minimax detector applies a clipping transformation 
to the nominal likelihood ratio function. The clipping
effect is achieved by shifting small portions of the probability
mass under each hypothesis to the tail sections where errors 
occur. This relatively minute shift of probability
mass can result in a significant degradation in test
performance.

We adopt here a minimax formulation of the robust
hypothesis testing problem of the same type as
\cite{Hub1,Hub2,Hub3}. The only difference is that the 
neighborhood where the actual observation probability 
density is located under each hypothesis is formed by
placing an upper bound on the relative entropy of the
actual density with respect to the nominal density.
To justify the choice of the relative entropy as a measure
of proximity between statistical models, observe that
Huber's work addresses primarily situations where 
statistical models are obtained directly from imperfect
data, possibly contaminated by outliers. However there
exists also situations where the densities employed in
hypothesis testing are model based, arising from 
physical considerations, possibly with a few unknown
parameters which are estimated from the data. In this
context, the relative entropy is a natural metric for
model mismatch, since it provides the underlying
metric for establishing the convergence of
the expectation-maximization method \cite{MK} of
mathematical statistics. In fact from a differential
geometric viewpoint, it is argued in \cite{AN} that
the relative entropy forms a natural `distance' between 
statistical models. More recently, in the context of
estimation and filtering it was shown in \cite{BJP,LN}
that minimax filters based on a relative entropy
tolerance take the form or risk-sensitive Wiener
or Kalman filters, which have well known robustness
properties. By selecting the relative entropy as a
measure of model mismatch, a risk-sensitive viewpoint
was also adopted recenty in \cite{HS} for developing
robust macroeconomic policies. Given relative entropy
neighborhoods of the nominal densities for the two 
hypotheses, it is easy to verify  that a
saddle point exists for the resulting minimax
hypothesis testing problem. To identify the saddle 
point, two assumptions are made. First as in
\cite{Hub1}, it is assumed that the nominal likelihood
ratio function (LR) is monotone increasing. Second,
it is required that the nominal densities under the
two hypotheses should be symmetric with respect to
each other. This allows the parametrization of the robust 
test and least-favorable densities in terms of a single 
parameter which can be selected uniquely so that the 
relative entropy tolerance is satisfied. The least-favorable
LR is expressed as a nonlinear transformation of the
nominal LR. But, unlike \cite{Hub1,Hub2,Hub3}, the
transformation is not a clipping transformation.
Instead, it attempts to drive the LR to a value as close
one as possible. The least-favorable densities
are divided into three segments. The extreme segments 
are scaled versions of the nominal densities, where the
scaling aims at shifting some probability mass to
tails where errors occur. But the middle segment
is a section of the ``mid-way density'' on the geodesic
linking the two nominal densities, where the mid-way
density is characterized by the property that 
it has the same relative entropy with respect to
each of the nominal densities. 

The robust hypothesis testing problem we consider
is also related to the worst-case noise detection
problem examined in \cite{MKV1,MKV2}, where 
given a binary communication system with additive
noise, with the actual noise density located
within a prespecified relative entropy bound of
the nominal noise density, it is required to find 
the ML detector for the worst-case noise in the 
neighborhood of the nominal noise. Thus the 
difference between the problem we consider and
\cite{MKV2} is that we allow the additive noise
statistics to be different under each hypothesis,
instead of forcing them to be the same. Finally,
it is worth noting that \cite{DJ} also examines 
robust hypothesis testing by using the relative
entropy as a mismatch metric between actual and
nominal densities, but it does so asymptotically
as the number of measurements becomes infinite,
so its results take a very different form.   

The paper is organized as follows. Section \ref{sec:formu}
describes the minimax hypothesis problem with a relative
entropy constraint. The saddle point of the problem
is characterized in Section \ref{sec:saddle}, and
examples are presented in Section \ref{sec:example}. 
Finally, Section \ref{sec:conc} gives some conclusions.
\vskip 2ex

\section{Problem Formulation}
\label{sec:formu}

Consider a binary hypothesis testing problem where under hypothesis 
$H_j$, with $j=0, \, 1$, the random observation $Y \in 
\mathbb{R}$ admits $f_j (y)$ as nominal probability density.
The actual density $g_j (y)$ of $Y$ under $H_j$ is not known
exactly and belongs to the neighborhood
\beq
{\cal F}_j = \{ g_j : \, D(g_j | f_j ) \leq \epsilon_j \}
\: , \label{2.1}
\eeq
where
\beq
D(g|f ) = \int_{-\infty}^{\infty} \ln \big( 
\frac{g(y)}{f(y)} \big) g(y) dy \label{2.2}
\eeq
denotes the Kullback-Leibler (KL) divergence or relative
entropy of probability densities $g(y)$ and $f(y)$.
Note that the KL divergence is not a true distance since it
is not symmetric, i.e., $D(g|f) \neq D(f|g)$, it does
not satisfy the triangle inequality, but $D(g|f) \geq 0$
with equality if and only if $g=f$. Also, since $x\ln(x)$
is a convex function for $x \geq 0$, $D(g|f)$ is convex
in $g$, which implies that neighborhood ${\cal F}_j$
is convex for $j=0 , \, 1$.

Let ${\cal D}$ denote the class of pointwise randomized
decision rules $\delta (y)$ such that if $Y=y$, we select
$H_1$ with probability $\delta(y)$ and $H_0$ with probability
$1-\delta$, where $0 \leq \delta(y) \leq 1$. Clearly
${\cal D}$ is convex, since if $\delta_1 (y)$ and $\delta_2 (y)$
are two decision rules of ${\cal D}$, then for $0 \leq
\alpha \leq 1$, 
\[
\delta (y) = \alpha \delta_1 (y) + (1-\alpha) \delta_2 (y)
\]
also belongs to ${\cal D}$.

Let 
\bea
P_F (\delta,g_0) &=& \int_{-\infty}^{\infty} \delta (y) g_0 (y) dy 
\label{2.3} \\
P_M (\delta,g_1) &=& \int_{-\infty}^{\infty} (1- \delta (y))
g_1 (y) dy \label{2.4}
\eea
denote respectively the probability of false alarm and
the probablity of a miss for decision rule $\delta \in {\cal D}$
when the densities of $Y$ under $H_0$ and $H_1$ are $g_0$
and $g_1$, respectively. Note that $P_F (\delta,g_0)$ is
separately linear in $\delta$ and $g_0$. Similarly 
$P_M (\delta,g_1)$ is separately linear in $\delta$ and $g_1$.
If we assume that the two hypotheses are equally likely,
the probability of error of $\delta \in {\cal D}$ is given by
\beq
P_E (\delta, g_0, g_1) = \frac{1}{2} [ P_F (\delta, g_0)
+ P_M (\delta, g_1)] \: . \label{2.5}
\eeq

We seek to solve the minimax problem 
\beq
\min_{\delta \in {\cal D}} \max_{(g_0 , g_1) \in {\cal F}_0
\times {\cal F}_1} P_E (\delta, g_0 , g_1) \label{2.6}
\eeq 
Note that $P_E (\delta,g_0,g_1)$ is linear and thus convex
in $\delta$. Similarly, it is linear and thus concave in 
$g_0$ and $g_1$. The set ${\cal F}_0 \times {\cal F}_1$
is convex and compact, ${\cal D}$ is convex and since
\[
||\delta||_{\infty} = \max_{y \in \mathbb{R}} \delta (y)
\]
for all $\delta \in {\cal D}$, ${\cal D}$ is compact with respect
to the infinity norm. So according to the Von Neumann minimax
theorem \cite[p. 319]{AE}, there exists a saddle point
$(\delta_{\mathrm{R}}, (g_0^{\mathrm{L}}, g_1^{\mathrm{L}}))$
for the minimax problem (\ref{2.6}). Here $\delta_{\mathrm{R}}$
is the robust/minimax test, whereas $g_0^{\mathrm{L}}$ and
$g_1^{\mathrm{L}}$ are the least favorable densities in
${\cal F}_0 \times {\cal F}_1$. The saddle point is characterized
by the property 
\beq
P_E (\delta, g_0^{\mathrm{L}}, g_1^{\mathrm{L}}) \geq
P_E (\delta_{\mathrm{R}}, g_0^{\mathrm{L}}, g_1^{\mathrm{L}}) 
\geq P_E (\delta_{\mathrm{R}}, g_0 , g_1 ) \label{2.7}
\eeq
for all $\delta \in {\cal D}$, $g_0 \in {\cal F}_0$ and
$g_1 \in {\cal F}_1$.

While it is nice to know that a saddle point exists,
exhibiting a test $\delta_{\mathrm{R}}$ and least favorable
densities $g_j^{\mathrm{L}}$, $j=0, \, 1$ satisfying
(\ref{2.7}) is a nontrivial task. Before doing so,
it is worth pointing out that the minimax problem
(\ref{2.6}) is of the same type as considered by Huber in
\cite{Hub1,Hub2,Hub3}. The only difference is that the 
neighborhoods ${\cal F}_j$ differ from those considered 
in \cite{Hub3} which included contamination models 
or proximity models based on the Kolmogorov metric
as special cases. The problem (\ref{2.6}) is also
closely related to the worst-case noise detection problem
considered in \cite{MKV2}, where for hypotheses
\bea
H_0 &:& Y= -1+N \nnl 
H_1 &:& Y= 1+N  \: , \label{2.8}
\eea
and a nominal probability density $f_N (n)$ for noise $N$, it
was desired to construct a minimum probability of error
detector for the least-favorable noise density $g_N (n)$
located in the KL ball specified by $D(g_N|f_N) \leq
\epsilon$. Thus the problem (\ref{2.6}) differs from the
one examined in \cite{MKV1,MKV2} by the fact that we allow 
the least-favorable noise distribution to be different under
hypotheses $H_0$ and $H_1$, instead of insisting they
should be the same.

\setcounter{equation}{0}
\section{Saddle Point Specification}
\label{sec:saddle}

The first inequality of the saddle point characterization 
(\ref{2.7}) indicates that the robust test $\delta_{\mathrm{R}}$
must be the optimum Bayesian test for the least-favorable
pair $(g_0^{\mathrm{L}} , g_1^{\mathrm{L}})$. So if
\beq
L_{\mathrm{L}} (y) = \frac{g_1^{\mathrm{L}} (y) }{
g_0^{\mathrm{L}} (y)} \label{3.1} 
\eeq
denotes the LR function for the pair
$(g_0^{\mathrm{L}}, g_1^{\mathrm{L}})$, we need to have
\beq
\delta_{\mathrm{R}} (y) = \left \{ \begin{array} {ccl}
1 & \mbox{for} & L_{\mathrm{L}} (y) > 1 \\
\mbox{arbitrary} & \mbox{for} & L_{\mathrm{L}} (y) = 1 \\
0 & \mbox{for} & L_{\mathrm{L}} (y) < 1 \: .
\end{array} \right. \label{3.2}
\eeq

Consider now the second inequality of (\ref{2.7}). Because of 
the form (\ref{2.5}) of $P_E (\delta , g_0 , g_1)$, it is equivalent
to
\bea
P_F (\delta_{\mathrm{R}}, g_0^{\mathrm{L}}) &\geq& 
P_F (\delta_{\mathrm{R}}, g_0) \nnl
P_M (\delta_{\mathrm{R}}, g_1^{\mathrm{L}}) &\geq& 
P_M (\delta_{\mathrm{R}}, g_1) \nn
\eea
for all $g_0$ and $g_1$ in ${\cal F}_0$ and ${\cal F}_1$,
respectively.

So, given $\delta_{\mathrm{R}}$, the least-favorable density
$g_0^{\mathrm{L}}$ is obtained by maximizing $P_F 
(\delta_{\mathrm{R}}, g_0)$ for all functions 
$g_0 \in {\cal F}_0$ such that
\beq
I(g_0) = \int_{-\infty}^{\infty} g_0 (y)dy = 1 \: . 
\label{3.3} 
\eeq
Since $P_F (\delta_{\mathrm{R}},g_0)$ is concave
in $g_0$ and the domain ${\cal F}_0$ is convex, 
the maximization can be accomplished by using the
method of Lagrange multipliers \cite[Chap. 5]{BNO}.
Consider the Lagrangian
\bea
\lefteqn{L(g_0,\lambda,\mu) = P_F (\delta_{\mathrm{R}},g_0 )
+ \lambda (\epsilon_0 - D(g_0|f_0)) + \mu (1-I(g_0))} \nnl
&=& \int_{-\infty}^{\infty} \big[ \delta_{\mathrm{R}} (y)
- \mu - \lambda \ln \big( \frac{g_0}{f_0} (y) \big) \big]
g_0 (y) dy + \lambda \epsilon_0 + \mu \: , \label{3.4}
\eea
where Lagrange multiplier $\lambda \geq 0$ is 
associated to the inequality constraint $D(g_0|f_0) \leq \epsilon_0$,
whereas multiplier $\mu$ corresponds to equality constraint
(\ref{3.3}). Note that the non-negativity constraint $g_0 (y)
\geq 0$ for the density function $g_0$ is not introduced 
explicitly, since the solution obtained below by maximizing
$L$ satisfies this constraint automatically. 

The Gateaux derivative \cite[p. 17]{BNO} of $L$ with respect
to $g_0$ in the direction of an arbitrary function $z$ is
given by
\bea
\nabla_{g_0 , z} L(g_0,\lambda,\mu) &=& \lim_{h \rightarrow 0}
\frac{1}{h} \big[ L(g_0+hz,\lambda,\mu) - L(g_0,\lambda,\mu)  
\big] \nnl
&=& \int_{-\infty}^{\infty} \big[ \delta_{\mathrm{R}} - (\lambda
+ \mu) -\lambda \ln \big( \frac{g_0}{f_0} \big) \big] zdy
\: , \label{3.5}
\eea
and since $z(y)$ is arbitrary, this implies
\beq
\delta_{\mathrm{R}} (y) - (\lambda + \mu) -\lambda \ln \big(
\frac{g_0}{f_0} \big) (y) = 0 \: . \label{3.6}
\eeq
In addition, the Karush-Kuhn-Tucker (KKT) condition
\beq
\lambda (\epsilon_0 -D(g_0|f_0)) = 0 \label{3.7}
\eeq
needs to be satisfied. Assume $\lambda >0$, so 
$D(g_0|f_0) = \epsilon_0$, i.e., $g_0$ is on the boundary
of ${\cal F}_0$. Then (\ref{3.6}) implies
\beq
g_0^{\mathrm{L}} (y) = \frac{1}{Z_0} \exp(\alpha_0 \delta_{\mathrm{R}} 
(y)) f_0 (y) \label{3.8} 
\eeq
with
\[
Z_0 \defn \exp(1 + \frac{\mu}{\lambda}) \hspace*{0.1in} , 
\hspace*{0.1in} \alpha_0 \defn \frac{1}{\lambda} \: .
\]
Note that since the nominal density $f_0 (y) \geq 0$
for all $y$, the least-favorable density $g_0^{\mathrm{L}} (y)$
specified by (\ref{3.8}) is also non-negative, so that
the non-negativity constraint on $g_0$ is satisfied 
automatically. Proceeding in a similar manner, 
we find that the least-favorable density under
$H_1$ can be expressed as
\beq
g_1^{\mathrm{L}} (y) = \frac{1}{Z_1} \exp(\alpha_1 (1-
\delta_{\mathrm{R}} (y) )) f_1 (y) \label{3.9}
\eeq
with $Z_1 >0$. 

Together, the expressions (\ref{3.2}) for $\delta_{\mathrm{R}}$
and (\ref{3.8})--(\ref{3.9}) for $(g_0^{\mathrm{L}},
g_1^{\mathrm{L}})$ provide some guidelines for guessing
a saddle point satisfying inequalities (\ref{2.7}). We
exhibit below a saddle point with the desired structure
under the following assumptions.
\vskip 2ex

\noindent
{\it Assumptions:} 

\begin{itemize}

\item[i)] The nominal likelihood ratio
\beq
L(y) = \frac{f_1 (y)}{f_0 (y)} \label{3.10}
\eeq
is a monotone increasing function of $y$. This implies
that $\ell = L(y)$ admits an inverse function $y = L^{-1}
(\ell)$.

\item[ii)] $f_0 (y)$ and $f_1 (y)$ admit the symmetry
\beq
f_1 (y) = f_0 (-y) \: . \label{3.11}
\eeq
This assumption implies
\[
L(-y) = \frac{f_1 (-y)}{f_0 (-y)} = \frac{f_0 (y)}{f_1 (y)}
= \frac{1}{L(y)} \: , 
\]
and thus $L(0) =1$.

\end{itemize}
\vskip 2ex

\noindent
{\it Remarks:}

\begin{itemize}

\item[a)] The motonicity assumption for $L(y)$ appears 
also in \cite{Hub3}. The symmetry condition (\ref{3.11})
has the effect of symmetrizing the KL divergence of
$f_0$ and $f_1$, since it ensures
\[
D(f_1|f_0) = D(f_0|f_1) \: .
\]
Furthermore, for $0 \leq u \leq 1$, if we consider the geodesic 
\[
f_u (y) = \frac{f_1^u (y) f_0^{1-u} (y)}{Z(u)}
\]
linking nominal densities $f_0$ and $f_1$, where
\[
Z(u) = \int_{-\infty}^{\infty} f_1^u (y) f_0^{1-u} (y) dy
\: ,
\]
the assumption ii) ensures that the density $f_{1/2}$ is located 
mid-way between $f_0$ and $f_1$ in terms of the KL divergence,
since
\[
D(f_{1/2}|f_0) = D(f_{1/2}|f_1) \: . 
\]
We refer the reader to \cite[Chap. 4.]{Cen} and \cite{AN} for a 
detailed discussion of the differential geometric structure of 
statistical models. 

\item[b)] For model (\ref{2.8}), the above assumptions are
satisfied if under both hypotheses $N$ admits a generalized 
Gaussian density 
\[
f_N (n) = a \exp(-|n/b|^{\alpha})
\]
with $\alpha >1$, where the constants $a$ and $b$ are adjusted 
to fix the variance of the distribution and normalize its total 
probability mass. The case $\alpha=2$ corresponds to a standard 
Gaussian distribution. On the other hand, if $N$ is Cauchy
distributed, it is easy to verify that 
\[
L(y) = \frac{f_N (y-1)}{f_N(y+1)} 
\]
is not monotone increasing so Assumption i) is not satisfied. 

\item[c)] The assumptions allow the consideration of
nonsymmetric noise distributions. For example, consider
model (\ref{2.8}) where under $H_1$, $N$ admits the
asymmetric Laplace density
\beq
f_L (n) = \left \{ \begin{array} {cc}
c \exp(-an) & n \geq 0 \\
c \exp(bn) & n \leq 0 
\end{array} \right. \label{3.12}
\eeq
with $b > a >0$ and $c = (a^{-1} + b^{-1})^{-1}$, and
under $H_0$, $N$ admits the flipped density $f_L (-n)$. 
Then
\[
f_1 (y) = f_L (y-1) \hspace*{0.1in} \mbox{and} \hspace*{0.1in}
f_0 (y) = f_L (-(y + 1))
\]
satisfy the symmetry condition (\ref{3.11}) and the 
log-likelihood ratio
\beq
\ln L(y) = \ln (f_1(y)/f_0 (y)) = \left \{ 
\begin{array} {cc} 
(b-a)y +(b+a) & y \geq 1 \\  
2by & -1 \leq y \leq 1 \\
(b-a)y -(b+a) &  y \leq -1
\end{array} \right. \label{3.13}
\eeq
is monotone increasing. Note that this property 
requires $b >a$, which ensures that the fat tails 
of $f_1(y)$ and $f_0(y)$ are located on the opposite
side of the location parameter of the competing hypothesis.
For example under $H_1$, the location parameter
(the constant additive term in (2.8)) is $1$
so the fat tail extends over $[1,\infty)$, 
which is on the opposite side of the location
parameter $-1$ of the competing hypothesis $H_0$. 

\end{itemize}
\vskip 2ex
 
We can now prove the following result.
\vskip 2ex

\begin{theo}
Assume that constants $\epsilon_j$ specifying
neighborhoods ${\cal F}_j$ with $j=0, \, 1$ are such that
$\epsilon_0 = \epsilon_1 = \epsilon$, where 
\beq
0 < \epsilon < D(f_{1/2}|f_0) \: . \label{3.14} 
\eeq
This requirement ensures that ${\cal F}_0$ and ${\cal F}_1$
do not intersect. Then under assumptions i)-ii) consider
the decision rule
\beq
\delta_{\mathrm{R}} (y) = \left \{ \begin{array}{cc}
1 & y > y_U \\[1ex]
\begin{displaystyle} \frac{1}{2} \big[ 1 + 
\frac{\ln L(y)}{\ln \ell_U} \big] \end{displaystyle} &
-y_U \leq y \leq y_U \\[1ex]
0 & y < -y_U  \: ,
\end{array} \right. \label{3.15}
\eeq
and the least-favorable pair
\bea
g_0^{\mathrm{L}} (y) &=& \left \{ \begin{array}{cc}
\ell_U f_0 (y)/Z(y_U) & y> y_U \\
\ell_U ^{1/2} f_1^{1/2} (y) f_0^{1/2} (y)/Z(y_U) & 
-y_U \leq y \leq y_U \\
f_0 (y)/Z(y_U) & y < -y_U 
\end{array} \right.  \label{3.16} \\
g_1^{\mathrm{L}} (y) &=& \left \{ \begin{array}{cc}
f_1 (y)/Z(y_U) & y> y_U \\
\ell_U^{1/2} f_1^{1/2} (y) f_0^{1/2} (y)/Z(y_U) & 
-y_U \leq y \leq y_U \\
\ell_U f_1 (y)/ Z(y_U) & y < -y_U 
\end{array} \right.  \label{3.17}
\eea
which are parametrized by $y_U >0$ and $\ell_U =
L(y_U) >1$. Here the normalizing constant $Z(y_U)$ is selected
such that
\beq
I(g_0^{\mathrm{L}}) = I(g_1^{\mathrm{L}}) = 1 \: .
\label{3.18}
\eeq
There exists a unique $y_U >0$ such that 
\beq
D(g_0^{\mathrm{L}}|f_0) = D(g_1^{\mathrm{L}}|f_1) =
\epsilon \: , \label{3.19}
\eeq
and the corresponding $\delta_{\mathrm{R}}$ and
densities $(g_0^{\mathrm{L}}, g_1^{\mathrm{L}})$
form a saddle point of minimax problem (\ref{2.6}).
\end{theo}
\vskip 2ex

Before proving the result, it is worth noting that
the least-favorable LR 
\beq
L_{\mathrm{L}} (y) = \left \{ \begin{array} {cc}
\begin{displaystyle} \frac{L(y)}{\ell_U} 
\end{displaystyle} >1 & y >y_U\\[1ex]
1 & -y_U \leq y \leq y_U \\
\ell_U L(y)  < 1  & y < -y_U 
\end{array} \right. \label{3.20}
\eeq
can be viewed as obtained by applying a nonlinearity
$q(\cdot)$ to the nominal likelihood ratio $L$. 
Specifically, we have
\beq
L_{\mathrm{L}} = q(L) = \left \{ \begin{array} {cc}
L/\ell_U & L > \ell_U \\
1 & \ell_U^{-1} \leq L \leq \ell_U \\
\ell_U L & L < \ell_U^{-1}
\end{array} \right. \label{3.21}
\eeq
where the nonlinearity $q(\cdot)$ is sketched in
\fig{nonlin} below. This nonlinearity is different
from the clipping transformation obtained by Huber
\cite{Hub1,Hub2,Hub3} which truncates high and low values
of the nominal likelihood ratio. Instead, the
transformation $q( \cdot )$ attempts to force the
transformed values $L_{\mathrm{L}}$ to be as close to
$1$ as possible, where a LR value $L_{\mathrm{L}} =1$
corresponds to a situation where observation $Y=y$ is 
uninformative in terms of making a decision between 
$H_1$ and $H_0$.  
\vskip 2ex

\bef[h]
\def\bdot {\fcir f:0 r:0.03 }

\def\Rtext #1{\bsegment
\textref h:L v:C \htext (0.1 0){#1}
\esegment}

\def\Ltext #1{\bsegment
\textref h:R v:C \htext (-0.1 0){#1}
\esegment}

\def\Ctext #1{\bsegment
\textref h:C v:C \htext (0 0){#1}
\esegment}

\def\Btext #1{\bsegment
\textref h:C v:T \htext (0 -0.1){#1}
\esegment}

\def\Ttext #1{\bsegment
\textref h:C v:B \htext (0 0.1){#1}
\esegment}

\begin{center}
\begin{texdraw}

\drawdim in
\linewd 0.01
\arrowheadsize l:0.08 w:0.05
\arrowheadtype t:F

\ravec (2.8 0) 
\Rtext{$L$}
\move (0.3 -0.3)
\ravec (0 1.8)
\Rtext{$q(L)$}
\move (0.3 0)
\rlvec (0.6 0.9)
\rlvec (0.75 0)
\rlvec (0.675 0.45)
\lpatt (0.05 0.05)
\move (0.3 0.9)
\Ltext{$1$}
\rlvec (0.6 0)
\rlvec (0 -0.9)
\Btext{$1/\ell_U$}
\move (1.65 0)
\Btext{$\ell_U$}
\rlvec (0 0.9) 
\move (0.15 -0.15)
\Ctext{$0$}
\move (1.85 1.25)
\Ctext{$L/\ell_U$}
\move (0.65 0.25)
\Ctext{$\ell_U L$}

\end{texdraw}
\end{center}
\caption{Nonlinearity $q(\cdot)$ relating the nominal and
least-favorable likelihood ratios.}
\label{nonlin}
\eef
\vskip 2ex

{\it Proof:} Observe first that since the
least-favorable LR is given by (\ref{3.20}), the
decision rule $\delta_{\mathrm{R}}$ specified
by (\ref{3.15}) has the form (\ref{3.2}). Note
that since $\ell_U^{-1} \leq L(y) \leq \ell_U$ for
$-y_U \leq y \leq y_U$, we have 
\[
-1 \leq \frac{\ln L (y)}{\ln \ell_U} \leq 1 
\]
for $-y_U \leq y \leq y_U$, which ensures $0 \leq
\delta_{\mathrm{R}} (y) \leq 1$ for $-y_U \leq y \leq y_U$.

Next, with $\delta_{\mathrm{R}}$ given by (\ref{3.15}).
it is easy to verify that the least favorable densities 
$g_0^{\mathrm{L}}$ and $g_1^{\mathrm{L}}$ given by 
(\ref{3.16}) and (\ref{3.17}) admit the forms (\ref{3.8})
and (\ref{3.9}) with $Z_0 = Z_1 = Z(y_U)$ and
\[
\alpha_0 = \alpha_1 = \ln \ell_U \: .
\]
To ensure that the normalization condition (\ref{3.18})
holds we only need to select
\[
Z(y_U) = \int_{-\infty}^{-y_U} f_0 (y) dy + \ell_U^{1/2}
\int_{-y_U}^{y_U} f_1^{1/2} (y) f_0^{1/2} (y) dy 
+ \ell_U \int_{y_U}^{\infty} f_0 (y) dy \: . 
\]
Then if $g_0^{\mathrm{L}} (\cdot |y_U)$ represents the function
(\ref{3.16}), where the parametrization by $y_U \geq 0$ is 
written explicitly, let
\bea
D(y_U) &\defn& D(g_0^{\mathrm{L}} (\cdot|y_U) |f_0) \nnl
&=& -\ln Z(y_U) + \frac{1}{Z(y_U)} \Big[
\ell_U \ln \ell_U \int_{y_U}^{\infty} f_0 (y)dy \nnl
&& \hspace*{1in} + \ell_U^{1/2} \ln \ell_U \int_0^{y_U} 
f_1^{1/2} (y) f_0^{1/2} (y) dy \Big] \label{3.22}
\eea
denote its KL divergence with respect to the nominal density
$f_0$. For $y_U =0$, we have $g_0^{\mathrm{L}} (\cdot|0) = f_0$,
so $D(0) =0$. Furthermore for $y_U = +\infty$, we have
$g_0^{\mathrm{L}} (\cdot| +\infty ) = f_{1/2}$, so 
$D(+ \infty) = D(f_{1/2}|f_0)$, where as noted earlier
the density $f_{1/2}$ represents the mid-way point on the 
geodesic linking $f_0$ to $f_1$.  

Taking the derivative of $D(y_U)$ with respect to $y_U$
gives
\bea
\frac{dD}{dy_U} &=& -  Z^{-1} (y_U) \frac{dZ}{dy_U} \nnl 
&& - Z^{-2} (y_U) \frac{dZ}{dy_U} \Big[ \ell_U \ln
\ell_U \int_{y_U}^{\infty} f_0 (y) dy + l_U^{1/2}
\ln \ell_U \int_{-y_U}^{y_U} f_1^{1/2} (y) f_0^{1/2} (y) dy 
\Big] \nnl
&& + Z^{-1} (y_U) \Big[ \frac{d~}{dy_U} (\ell_U \ln
\ell_U ) \int_{y_U}^{\infty} f_0 (y) dy + \frac{d~}{dy_U}
( \ell_U^{1/2} \ln \ell_U ) \int_0^{y_U} f_1^{1/2} (y) 
f_0^{1/2} (y) dy \Big] \nnl
&=&  \frac{N(y_U)}{Z^2 (y_U)} \frac{dL}{dy_U} \: , \label{3.23} 
\eea
where 
\bea
&& N(y_U) = \ln \ell_U \int_{y_U}^{\infty} f_0 (y) dy
\int_{y_U}^{\infty} f_1 (y) dy \nnl
&& \hspace*{0.2in} + \frac{1}{2} \ln \ell_U 
\Big[ \ell_U^{1/2} \int_0^{y_U} f_1^{1/2} (y) f_0^{1/2}
(y) dy + \int_{y_U}^{\infty} \big( f_0 (y) + \frac{f_1 (y)}{
\ell_U} \big)dy \Big] > 0 \nn
\eea
for $y_U >0$. Since $L(y)$ is monotone increasing, we have $dL/dy_U
> 0$ in (\ref{3.23}), so $dD/dy_U >0$. Consequently,
$D(y_U)$ is monotone increasing from $D(0) =0$ for $y_U=0$
to $D(f_{1/2}|f_0)$ for $y_U = \infty$. Accordingly,
given $\epsilon$ satisfying (\ref{3.14}), there exists
a unique $y_U$ such that $D(y_U) = \epsilon$. For this
choice of $y_U$, the least favorable densities
$g_0^{\mathrm{L}}$ and $g_1^{\mathrm{L}}$ satisfy KKT
condition (\ref{3.9}), so the second inequality of
(\ref{2.8}) is satisfied, and $\delta_{\mathrm{R}}$
together with $(g_0^{\mathrm{L}}, g_1^{\mathrm{L}})$
form the desired saddle point. \hfill $\Box$  
\vskip 2ex 

\noindent
{\it Worst case test performance:} By taking into account
the symmetries
\bea
1 - \delta_{\mathrm{R}} (y) &=& \delta_{\mathrm{R}} (-y) \nnl
g_1^{\mathrm{L}} (y) &=& g_0^{\mathrm{L}} (-y) \label{3.24} 
\eea
of the robust test and least favorable densities, which are
a consequence of the symmetry assumption (\ref{3.11}), we find
that the worst-case probabilities of false alarm and of a miss
for test $\delta_{\mathrm{R}}$ satisfy
\[
P_F (\delta_{\mathrm{R}},g_0^{\mathrm{L}}) =
P_M (\delta_{\mathrm{R}},g_1^{\mathrm{L}}) =
P_E (\delta_{\mathrm{R}},g_0^{\mathrm{L}}, g_1^{\mathrm{L}}) 
\: , 
\]
where
\bea
P_F (\delta_{\mathrm{R}},g_0^{\mathrm{L}}) &=&
\int_{-y_U}^{y_U} \delta_{\mathrm{R}} (y) g_0^{\mathrm{L}} (y) dy
+ Z^{-1} (y_U) \ell_U \int_{y_U}^{\infty} f_0 (y) dy \nnl
&=& Z^{-1} (y_U) \Big[ \ell_U^{1/2} \int_0^{y_U} f_1^{1/2} (y)
f_0^{1/2} (y) dy + \ell_U \int_{y_U}^{\infty} f_0 (y) dy \Big]
\: . \label{3.25} 
\eea

\setcounter{equation}{0}
\section{Examples}
\label{sec:example}

\noindent
{\it Example 1:} Consider the case where under $H_0$ and $H_1$, $Y$ 
admits the nominal distributions
\bea
f_0 (y) &=& \frac{1}{(2\pi \sigma^2)^{1/2}} \exp \big(
-\frac{(y+1)^2}{2\sigma^2} \big) \nnl
f_1 (y) &=& \frac{1}{(2\pi \sigma^2)^{1/2}} \exp \big(
-\frac{(y-1)^2}{2\sigma^2} \big) \: . \label{4.1}
\eea
This corresponds to a model of the form (\ref{2.8}) 
where the additive noise $N$ has a $N(0,\sigma^2)$
nominal distribution. The signal to noise ratio (SNR)
for this detection problem is $\mathrm{SNR} = 1/\sigma^2$.
The likelihood ratio
\[
L(y) = \frac{f_1 (y)}{f_0 (y)} = \exp \big( \frac{2y}{
\sigma^2} \big) 
\]
is clearly monotone increasing, and the nominal
densities $f_j(y)$, $j=0, \, 1$ admit the symmetry 
(\ref{3.11}), so the assumptions of Theorem 1 are satisfied.
In this case, it is interesting to note that the
mid-way density
\[
f_{1/2} (y) = \frac{f_1^{1/2} (y)f_0^{1/2} (y)}{Z(1/2)}
= \frac{1}{(2\pi\sigma^2)^{1/2}} \exp \big (
- \frac{y^2}{2\sigma^2} \big) \: ,
\]
is $N(0,\sigma^2)$ distributed, which makes sense 
since $f_0$ and $f_1$ have opposite means $\mp 1$ but the
same variance $\sigma^2$.

If we consider the parametrization (\ref{3.16}) 
of the least favorable density $g_0^{\mathrm{L}} (y)$,
we find that it is continuous and formed by three  
segments. Over $(-\infty, -y_U)$, $g_0^{\mathrm{L}}$ is an
attenuated version of the nominal $N(-1,\sigma^2)$ density.
Over $[-y_U, y_U]$, it is a scaled version of the
mid-way $N(0,\sigma^2)$ density, and for $(y_U,\infty)$
it is an amplified  version of the nominal $N(-1,
\sigma^2)$ density. Thus $g_0^{\mathrm{L}}$ can be viewed
as obtained from the nominal density $f_0$ by shifting
a portion of its probability mass to the middle segment
where $g_0^{\mathrm{L}}$ and $g_1^{\mathrm{L}}$ are
equal, and to the right tail where hypothesis $H_1$ is
selected, which has the effect of increasing the probability 
of false alarm.
\vskip 2ex

\bef[h]
\centering
\includegraphics[width = 3.5in,height =3in]{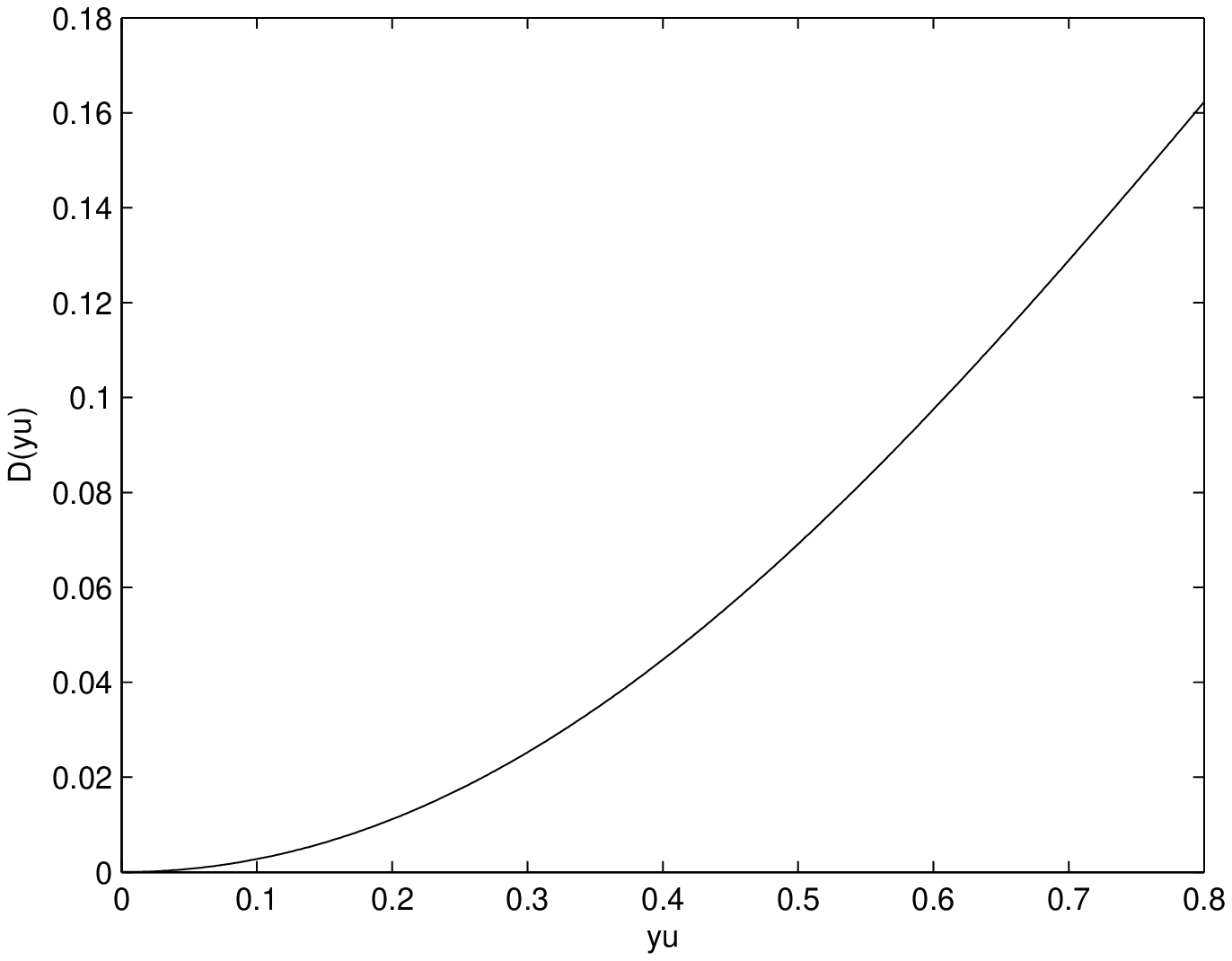}
\caption{Plot of function $D(y_U)$ for $0$dB SNR.}
\label{kldiv}
\eef
\vskip 2ex

To illustrate the construction of $g_0^{\mathrm{L}} (y)$,
let the relative entropy tolerance be $\epsilon = 0.1$.
Then for a nominal SNR equal to $0$dB ($\sigma =1$), the 
function $D(y_U)$ measuring the KL divergence of
$g_0^{\mathrm{L}} (\cdot|y_U)$ with respect to $f_0$
is plotted in \fig{kldiv}. As expected, it is monotone
increasing and attains the desired tolerance value
$\epsilon = 0.1$ for $y_U = 0.6080$. The least-favorable
density $g_0^{\mathrm{L}} (y)$ is plotted together with the
nominal density $f_0 (y)$ in part a) of \fig{lfgaus}.
\vskip 2ex

\bef[htb!]
\centering
\includegraphics[width = 3.5in,height =3in]{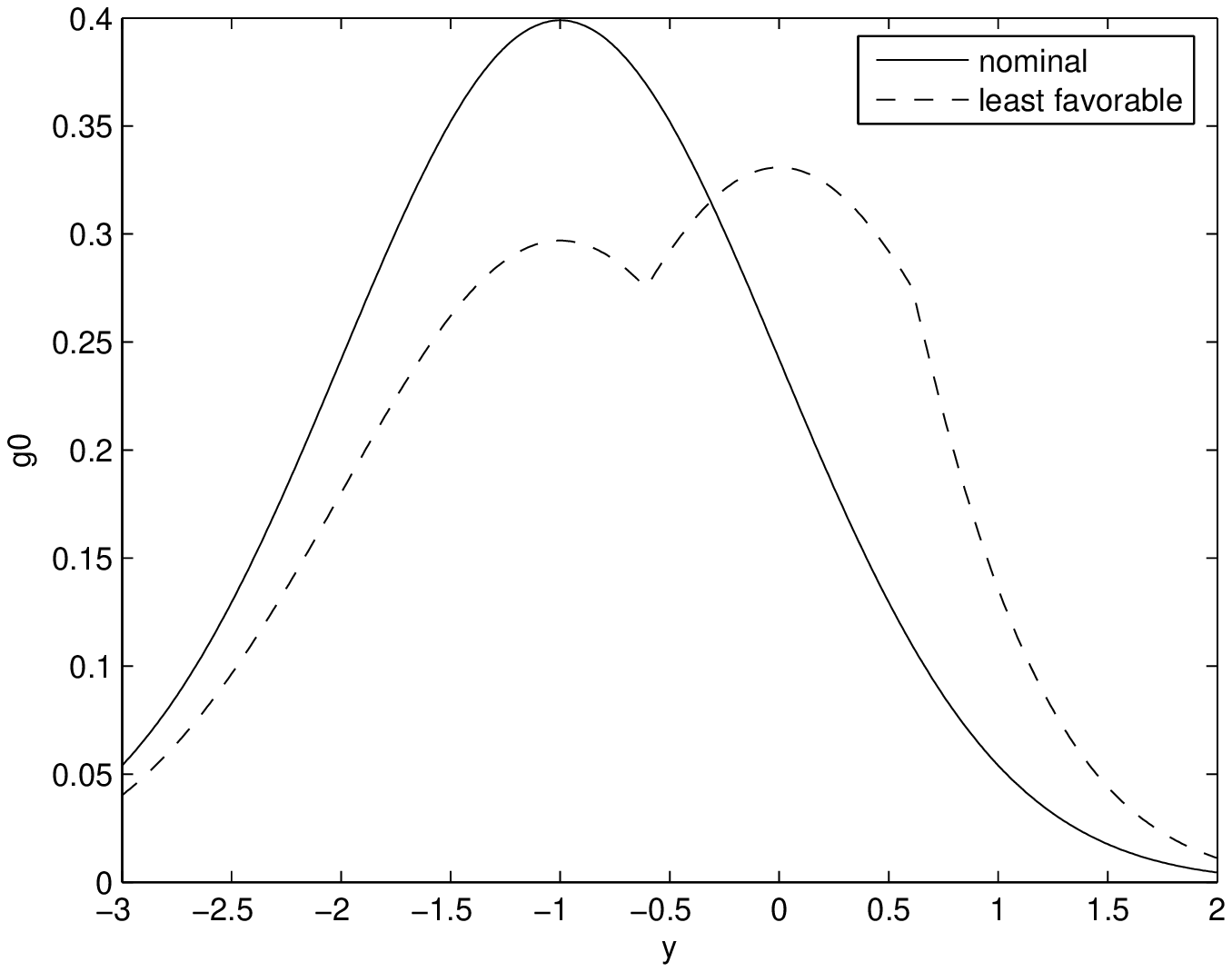}
\vskip 1ex
(a) 
\vskip 1ex
\includegraphics[width = 3.5in,height =3in]{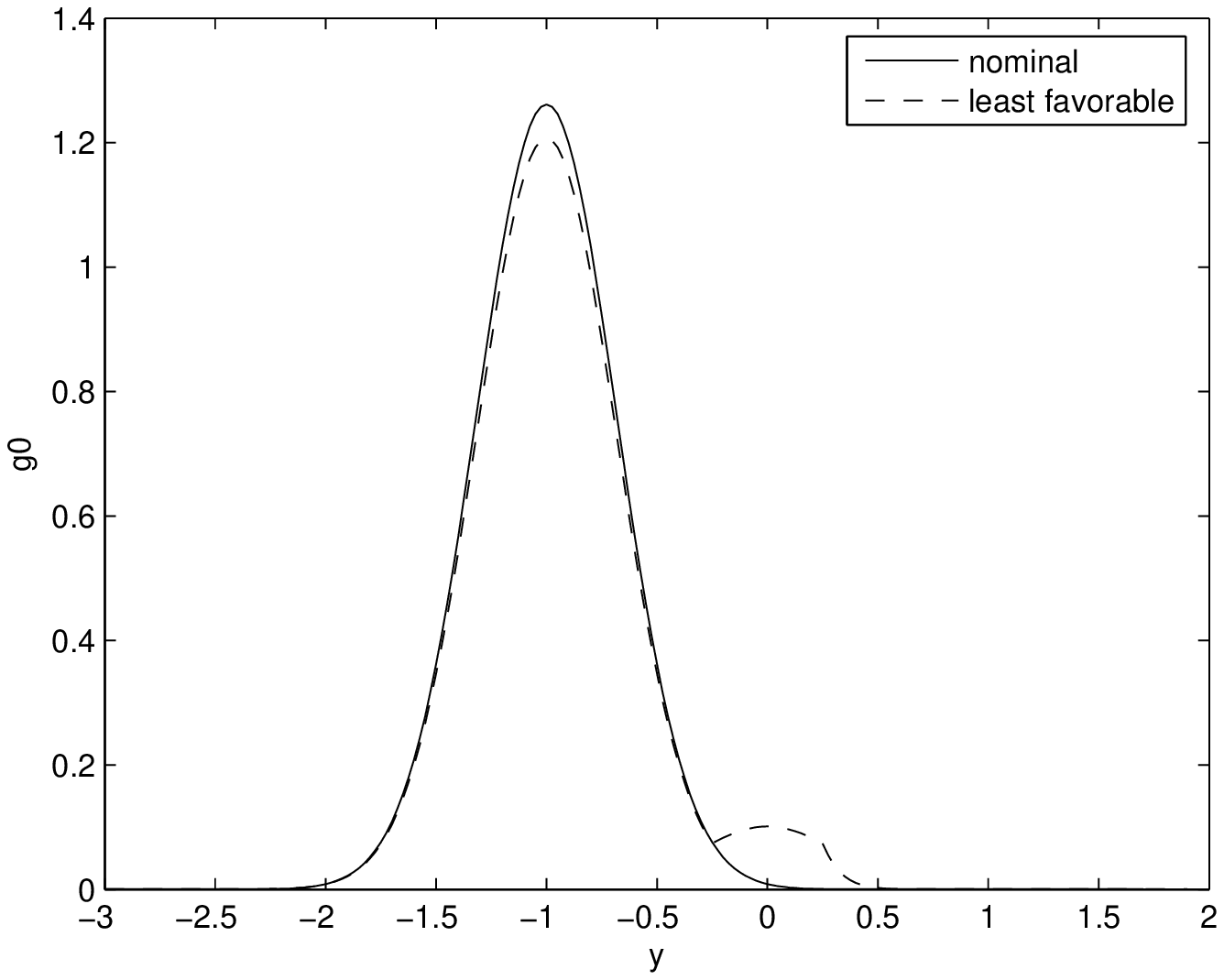}
\vskip 1ex
(b) 
\caption{Least favorable density $g_0^{\mathrm{L}} (y)$ for
a tolerance $\epsilon = 0.1$ and a) SNR =0dB, b) SNR = 10dB.}
\label{lfgaus}
\eef
\vskip 2ex

\noindent
The three segments of the density described earlier are
clearly in evidence in this plot. Note however that as
the SNR increases, the middle segment shrinks. For example,
the least-favorable density for a SNR value of $10$dB is
shown in part b) of \fig{lfgaus}. Although the KL
tolerance $\epsilon = 0.1$ is the same as in part a),
the deviation of $g_0^{\mathrm{L}}$ away from $f_0$
is much smaller than for a SNR value of 0dB. Note also
that $g_0^{\mathrm{L}}$ is not symmetric about $-1$ 
since a fraction of the probability mass has been 
transferred from the left tail to the right tail in
the direction of the location parameter $1$ of the competing
hypothesis $H_1$. Similarly the least favorable distribution
$g_1^{\mathrm{L}}(y) = g_0^{\mathrm{L}} (-y)$ transfers
a portion of its probability mass from its right tail
to its left tail. In terms of model (\ref{2.8}),
this means that the least favorable densities of the
noise $N$ are different under $H_0$ and $H_1$, since one
tilts rightward while the other tilts leftward. In
contrast, \cite{MKV2} requires that the least-favorable 
noise should be the same under both hypotheses. For
the above example with $\epsilon =0.1$ and $10$dB SNR,
the least favorable noise density is plotted in the
SouthWest corner of Figure 3 of \cite{MKV2}. It is  
symmetric and thus differs from the least-favorable
densities obtained here.

Finally, for $\epsilon = 0.01$ and $0.1$, and for SNR 
values between $0$ and $15$dB, the worst-case performance 
of the robust test $\delta_{\mathrm{R}}$ given by (\ref{3.25})
is compared in \fig{perf} with the probability  of error
$P_E = Q( \mathrm{SNR}^{1/2})$ of the maximum likelihood 
detector for nominal densities (\ref{4.1}). As indicated by
the figure, the loss of performance is rather spectacular. 
Of course, since this performance represents a worst case
situation, it is not truly indicative of the degradation
incurred for more benign choices of densities $g_j$ in
${\cal F}_j$ with $j=0,\, 1$. 
\vskip 2ex

\bef[htb!]
\centering
\includegraphics[width = 3.5in,height =3in]{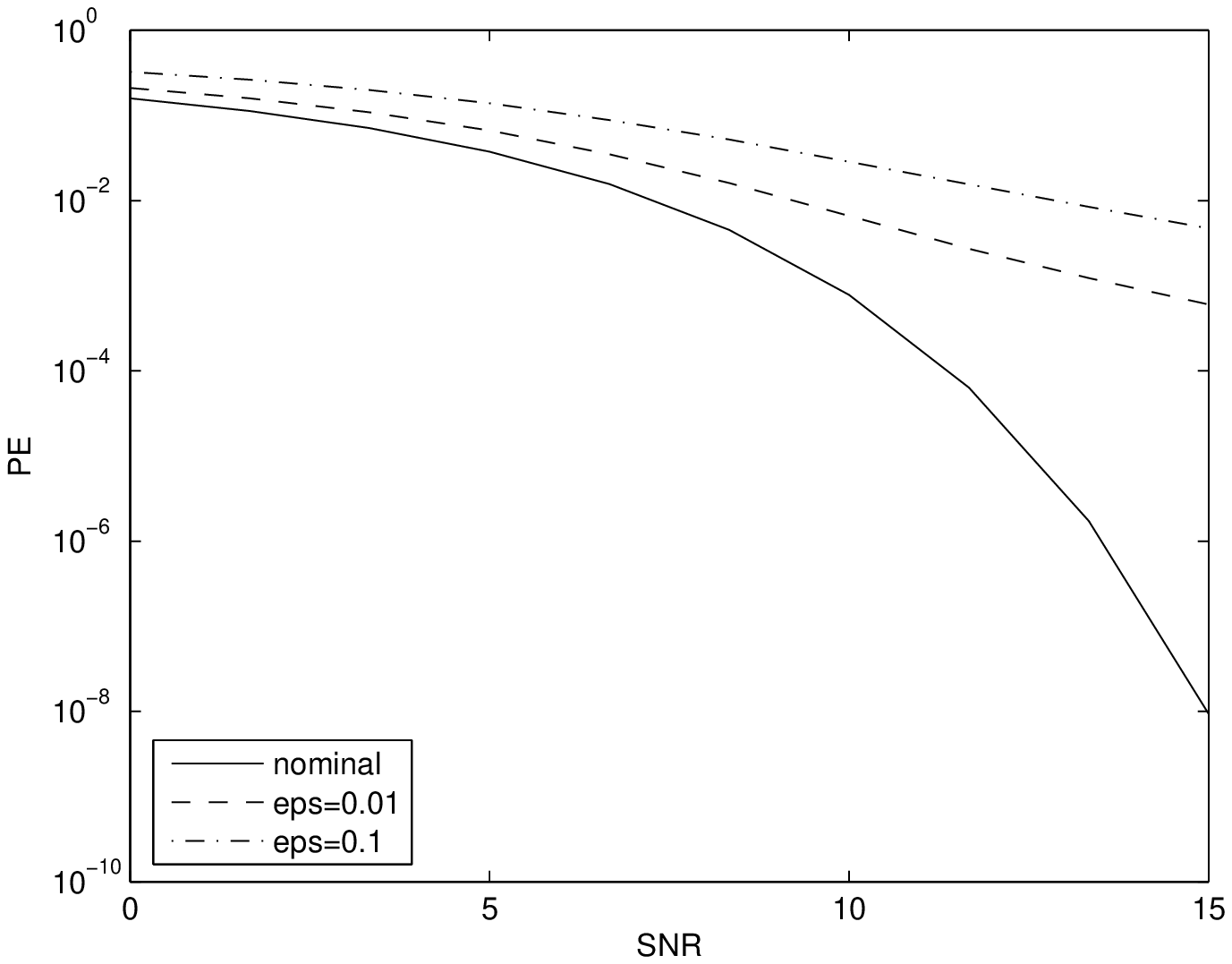}
\caption{Comparison of the worst case probability  of error
of test $\delta_{\mathrm{R}}$ for $\epsilon = 0.01$ and 
$\epsilon = 0.1$ with the ML probability of error
for the nominal model.} 
\label{perf}
\eef
\vskip 2ex

\noindent
{\it Example 2:} Consider model (\ref{2.8}) where under $H_1$
$N$ admits the asymmetric Laplace density $f_L (n)$ given by 
(\ref{3.12}) with $b>a$ and under $H_0$, $N$ admits the flipped 
density $f_L (-n)$. Then the densities
\[
f_1 (y) = f_L (y-1) \hspace*{0.1in} , \hspace*{0.1in}
f_0 (y) = f_L (-(y+1)) 
\]  
satisfy the symmetry condition (\ref{3.11}), and as
indicated by (\ref{3.13}), the likelihood ratio $L(y)$
is monotone increasing. In this case, the half-way density 
\bea
f_{1/2} (y) &=& \frac{f_0^{1/2} (y)f_1^{1/2} (y)}{Z(1/2)} \nnl
&=& \left \{ \begin{array} {cc}
c \exp(-b)/Z(1/2) & -1 \leq y \leq 1 \\[1ex]
c\exp \big( -\frac{a+b}{2}|y| + \frac{a-b}{2} \big)/Z(1/2) & |y| \geq 1
\end{array} \right. \nn
\eea  
with
\[
Z(1/2) = 2c \exp(-b) \big( 1 + \frac{2}{a+b} \big)
\]
is constant for $-1 \leq y \leq 1$ and has a symmetrized 
exponential decay rate for its two tails. For
$y_U <1$, the parametrization (\ref{3.16}) of the 
least favorable density $g_0^{\mathrm{L}}$ indicates
that over segments $(-\infty ,-y_U)$ and $(y_U,\infty)$
it is proportional to $f_0$, but over $[-y_U,y_U]$ it
is constant since $f_{1/2}$ is constant. 

To illustrate this feature the nominal and least favorable
densities are plotted in \fig{lflap} for $a=2$, $b=4$,
and $\epsilon =0.1$. For this choice of parameters $y_U = 0.3640$.  
\vskip 2ex

\bef[htb!]
\centering
\includegraphics[width = 3.5in,height =3in]{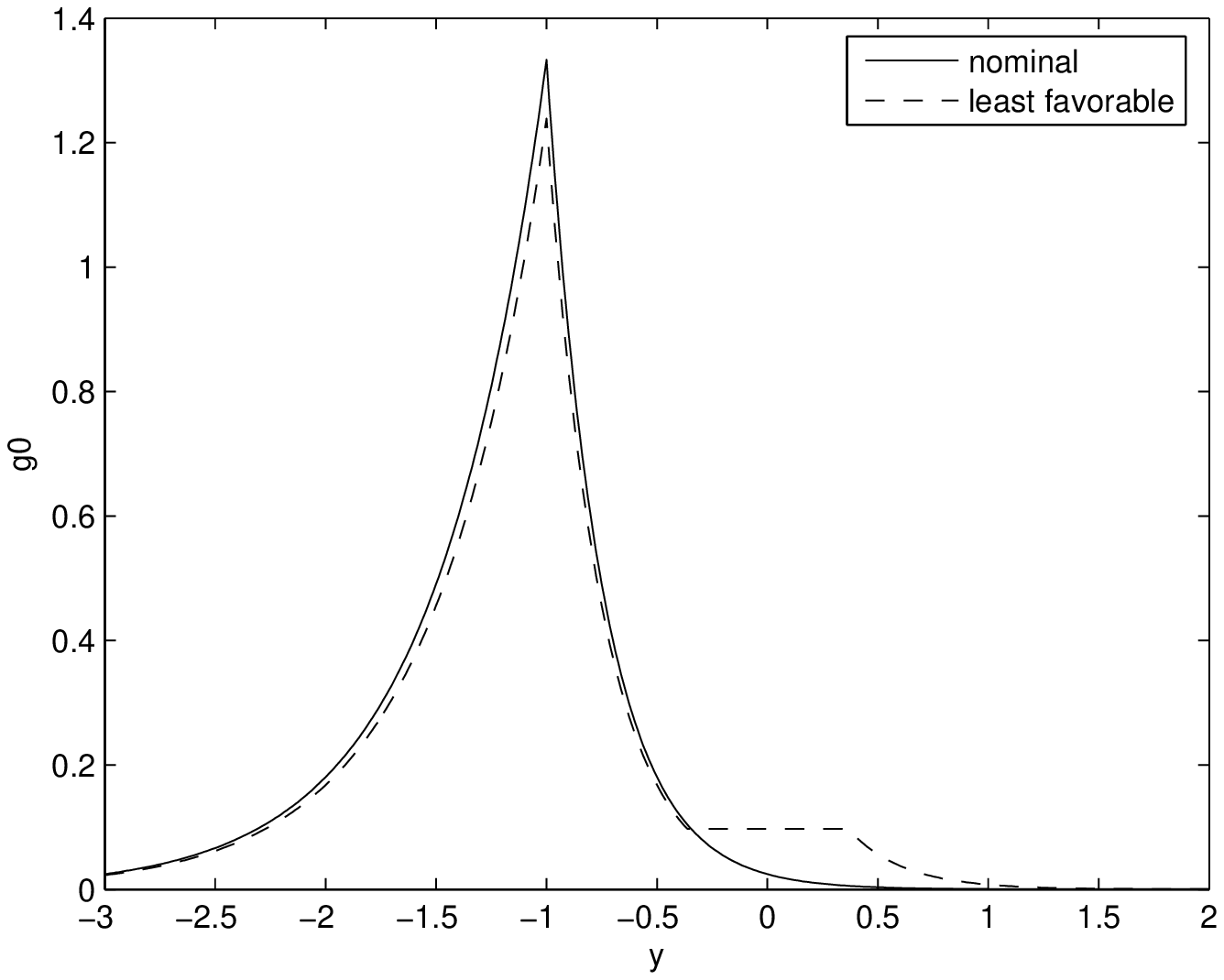}
\caption{Nominal asymmetric Laplace density $f_0$ and least 
favorable density $g_0^{\mathrm{L}}$ for $a=2$, $b=4$ and 
tolerance $\epsilon = 0.1$.}
\label{lflap}
\eef
\vskip 2ex

\section{Conclusion}
\label{sec:conc}

A minimax hypothesis testing procedure has been derived
for a binary hypothesis testing problem where
the actual observation density under each hypothesis is 
required to be within a fixed KL ball centered about
the nominal density. The robust test applies a nonlinear
transformation which flattens the nominal LR in the
vicinity of $L=1$. The least-favorable densities
include three segments where, quite interestingly,
the middle segment is formed by a section of the density
located mid-way on the geodesic linking the nominal densities
under the two hypotheses.

The results were derived under a motonicity condition
for the LR as well as a symmetry condition for the
two hypotheses. While the first condition is benign
and appears in Huber's work \cite{Hub1,Hub2,Hub3},
it would be desirable to remove the symmetry condition
(\ref{3.11}), since this would open the way to the study
of more general robust signal detection problems of the 
type discussed in \cite{KP}.

\end{document}